\documentclass[11pt,twoside]{article}

\usepackage{asp2006}
\usepackage{epsf}
\usepackage{lscape}

\newcommand{\more}{\raisebox{-1.1mm}{$\stackrel{>}{\sim}$}}

\begin{document}
\title{AGB stars in extragalactic systems}   
\author{M.A.T. Groenewegen}   
\affil{Instituut voor Sterrenkunde, Celestijnenlaan 200 D, B-3001 Heverlee, Belgium (groen@ster.kuleuven.be)}

\begin{abstract} 

I will be reviewing three methods to identify late-type giants in
extragalactic systems, based on the main characteristics of AGB stars
(they are infrared bright, variable, and have spectral peculiarities).

\end{abstract}
\vspace{-10mm}



\section{Introduction}  

All main-sequence stars born with masses below $\la$ 8 $M_{\odot}$
have or will go through the evolutionary phase called Asymptotic Giant
Branch (AGB). The lower limit in initial mass is set by the age of the
Galactic Disc, the upper limit is set by the mass where carbon can be
ignited in the stellar core (but see the presentation by Siess on
super-AGB stars in this volume). The AGB is the final phase where
intermediate-mass stars have nuclear burning in the form of alternate
Hydrogen and Helium shell burning, before they cross the
Herzsprung-Russell diagram to become Planetary Nebulae and then White
Dwarfs.

A summary of the interior structure and stellar evolution up to and on
the AGB, and including the post-AGB phase, can be found in the recent
textbook ``Asymptotic Giant Branch Stars'' (Habing \& Olofsson 2004).

AGB stars are luminous ([$\sim$0.1 -- a few] 10$^4$ $L_{\odot}$) and
cool, with effective temperatures in the range 3850 to $\sim$2500 K
(for M0 to M10 giants, e.g. Fluks et al. 1994). From this it follows
that AGB stars are big (up to a few hundred $R_{\odot}$), and
combining this with the classical pulsation equation $P = 0.038 \;
R^{1.5} \; M^{-0.5}$ it follows that any fundamental mode radial
pulsations that would occur would have periods between tens and
hundreds of days.

Equally important, and typical for the AGB, are the chemical
peculiarities that occur during this evolutionary phase (see Chapter 2
in the aforementioned book). Depending in a complex way on initial mass,
metallicity, mass-loss, mixing and burning in the envelope [hot
bottom-burning], an AGB star may go through several third dredge-up
events whereby mainly carbon, nitrogen, oxygen and $s$-process
elements are mixed ultimately into the stellar photosphere. Depending
on the C/O-ratio different molecules form in the cool atmospheres
(VO, TiO, C$_2$, CN) and a star can be classified as M-star (C/O $\la$
0.95), S-star (0.95 $\la$ C/O $<$ 1.0) or C-star (C/O $\ge$ 1.0). 
Intermediate classes (MS, SC) also exist.

The low effective temperatures already make AGB stars redder than all
their Main-Sequence progenitors. In addition, the formation of
different molecules depending on chemical type makes that the infrared
colours of M- and C-stars are different, as will be discussed later.

Furthermore, for spectral types later than $\sim$M4-M5 (e.g. Glass \&
Schultheis 2002) the region close to the star has the right
combination of temperature and density for dust grains to form. Dust
absorbs efficiently in the optical and radiates in the infra-red. This
implies that AGB stars surrounded by dust shells are even redder.

In the present review I will only discuss the most recent results. For
earlier reviews covering AGB stars in Local Group (LG) galaxies see,
Azzopardi (1999), and Groenewegen (1999, 2002, 2006a,b).
\vspace{-5mm}

\section{Searches for AGB stars}

\subsection{ Near- and Mid-infrared studies }

As AGB stars are cool and many lose mass, infra-red colours are a
natural way to search for them. A disadvantage is that only candidates
may be identified, although the $(J-K)$ colour is often used to
discriminate M and C-stars (see the discussion later).
An advantage is that very red colours trace a different population of
extreme mass loss and likely of higher initial mass, or stars at the
very end of the AGB.

As the 2MASS $JHK$ survey (Cutri et al. 2003) was an all-sky survey that went
reasonably deep ($K \approx 15$), it had the potential of discovering
AGB stars in nearby LG galaxies. Quite a few studies have appeared
that used 2MASS data: 
Demers et al. (2002) correlated spectroscopically confirmed C-stars in
the Magellanic Clouds (MCs) with 2MASS and used this to propose 26
C-star candidates in the Fornax DSph.
Cioni et al. (2003) studied the spatial C/M ratio over the MCs (also
using DENIS data) to infer the distribution in [Fe/H].
Tsalmantza et al. (2006) used virtual observatory tools to identify
luminous ($M_{\rm bol} < -6.0$) AGB stars with $(J-K) > 1.5$ and
$(H-K)> 0.4$ in the MCs, M31 and M33.
Groenewegen (2006a) selected 2MASS sources in LG galaxies within 1 Mpc
(but excluding MCs, M31, M32, M33), and retaining objects with $(J-K)_0 >
1.22$, appropriate $M_{\rm K}$-range for AGB stars, errors in $J,K <
0.12$ and excluding known objects using the SIMBAD database.

In the last couple of years more studies appeared that used ground-based IR
instrumentation sometimes combined with an optical colour.
Cioni \& Habing (2005a) studied a field in Draco of 40\arcmin\ x30\arcmin\ in $IJK$, and
Cioni \& Habing (2005b) studied a field in NGC 6822 of 20\arcmin\ x20\arcmin\ in $IJK$.
At this conference she also presented preliminary results on M33. 
In her papers she selects O-rich and C-rich AGB stars from their infrared colours and then 
studies the spatially resolved C/M ratio.
Kang et al. (2006) studied a smaller field of 6.3\arcmin\ x 3.6\arcmin\ in $giJHK$ in NGC 6822.
Since this galaxy was observed using the narrow-band filter technique
(see later) by Letarte et al. (2002) they could identify known carbon stars in various
colour-colour diagrams.
The M31 companions NGC 147, 185, 205 have been studied by Davidge (2005), Kang et al. (2005) 
and Sohn et al. (2006).
Finally, Rejkuba et al. (2006, and this conference) present NIR data on dwarf ellipticals in Cen A.

\smallskip
Turning to the mid-IR, the {\it Spitzer Space Telescope} has a great
discovery potential for luminous and especially mass-losing AGB stars
in LG galaxies. First results have been published from the S$^3$MC
survey of the SMC (Bolatto et al. 2006, program ID 3316), the SAGE
survey of the LMC (Meixner et al. 2006, Blum et al. 2006, program ID
20203), and for M31 (Barmby et al. 2006, program ID 3126).
Smaller LG galaxies will also be surveyed in both IRAC (P.I. R.~Gehrz,
program ID 128) and MIPS (P.I. E~Skillman, program ID 20425).

First results on IRAC observations of WLM were presented by Dale
Jackson et al. at this conference.  As WLM has been surveyed using the
narrow-band filter system (Battinelli \& Demers 2004) they could
compare the number of detections.  Interestingly, they recovered 90\%
of the known C-stars in the IRAC colours, but this represented only of
order 20\% of the entire AGB population.  Especially stars with
$[3.6-4.5] \more 0.3$ were lacking among the known C-stars (and one
would expect them to be C-stars at the low metallicity of WLM).

\begin{figure}[!ht]

\centering

\begin{minipage}{0.50\textwidth}
\resizebox{\hsize}{!}{\includegraphics{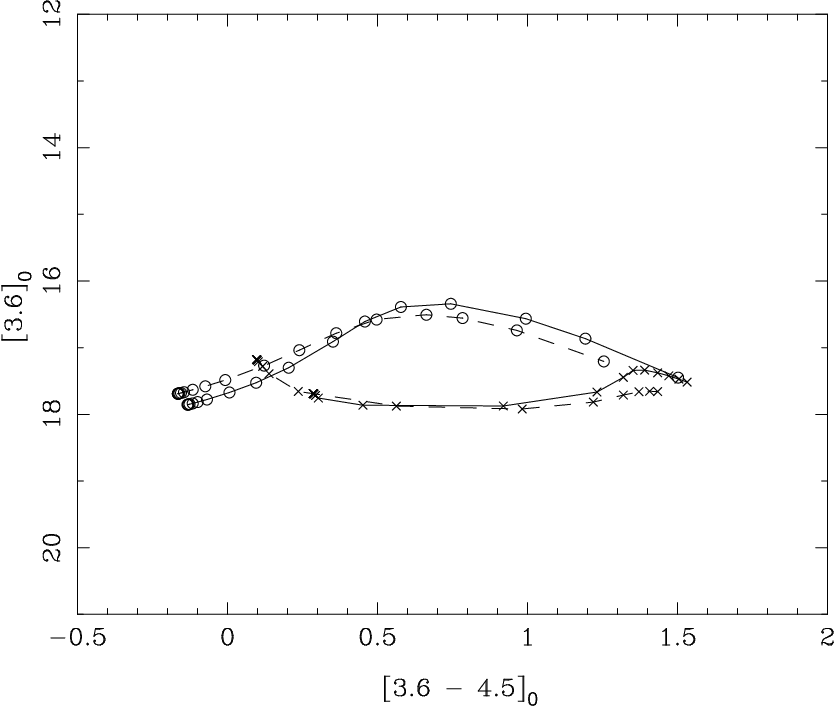}}
\end{minipage}


\caption{IRAC colour-magnitude diagram for models with 3000 L$_{\odot}$ at 932 Kpc (representative for WLM). 
The following sequences of increasing mass loss are shown: 
Carbon-rich AGB star with $T_{\rm eff}$ = 3600 K, and 85\% AMC + 15\% SiC (circles \& solid line), 
Carbon-rich post-AGB star with $T_{\rm eff}$ = 2650 K, and 100\% AMC (crosses \& solid line), 
Oxygen-rich AGB star with $T_{\rm eff}$ = 3297 K, and 60\% silicate + 40\% AlOx (circles \& dashed line), 
and Oxygen-rich post-AGB star with $T_{\rm eff}$ = 2500 K, and 100\% silicate (crosses \& dashed line). 
From Groenewegen (2006c).  }
\label{groen-WLM}
\end{figure}

To aid in identifying AGB stars in the non-standard IRAC and MIPS
filters, Groenewegen (2006c) has presented tracks for AGB stars in
various optical, near- and mid-infrared colours as a function of mass
loss rate and for both O- and C-stars. An example is shown in Fig.~\ref{groen-WLM}. 
Note that the fluxes and mass loss rates listed in that paper are for
a particular luminosity and distance, and scaling relations need to be
applied (as explained in the paper) to compute fluxes and mass loss
rates for other values of luminosity and distance !

\subsection{ Variability }

Variability is another main characteristic of AGB stars. Mira
variables have peak-to-peak amplitudes of larger than 2.5, 0.9, and
0.4 mag in $V,I,K$, respectively, and are therefore easy to find.  The
disadvantage is that pulsation amplitude and period are not sufficient
to discriminate O-rich from C-rich objects, and that the observations
are (observing) time demanding. An advantage is that Miras follow a
tight period-luminosity relation and hence (relative) distances can be derived.

\begin{figure}[!t]
\begin{minipage}{0.49\textwidth}
\resizebox{\hsize}{!}{\includegraphics{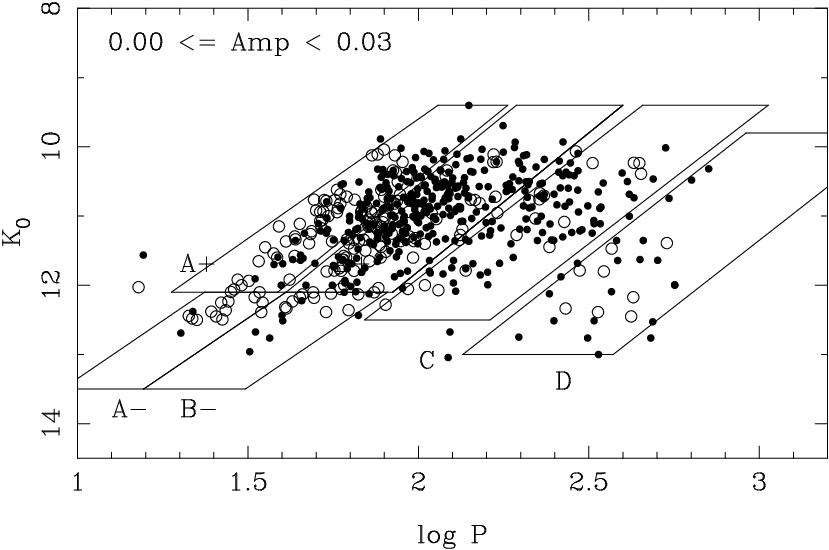}}
\end{minipage}
\hfill
\begin{minipage}{0.49\textwidth}
\resizebox{\hsize}{!}{\includegraphics{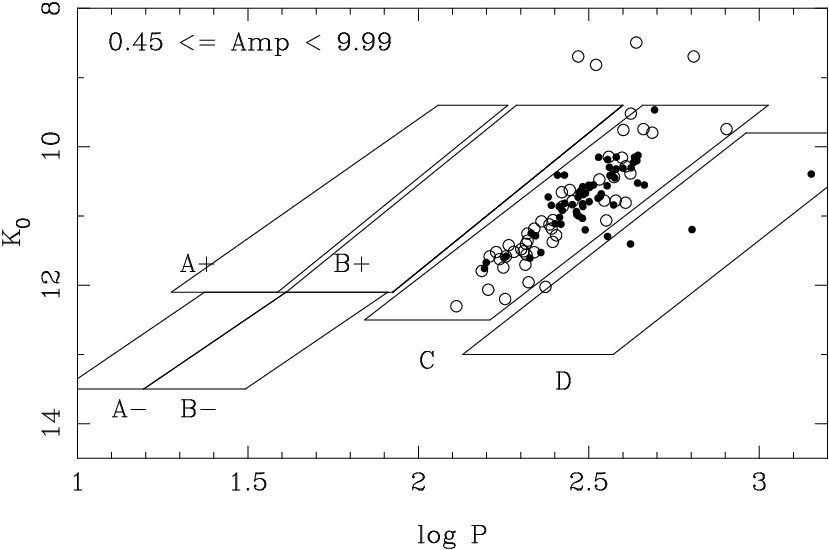}}
\end{minipage}
\caption{
$K$-band $PL$-relation for the LMC. Panels indicate selection on
$I$-band amplitude as indicated in the upper left corners.  Carbon
stars are indicated by filled circles, M- and S-stars by open
circles. Boxes related to the ``ABCD'' (Ita et al. 2004) sequences are
indicated. From Groenewegen (2004).
}
\label{groen-PL-MC}
\vspace{-3mm}
\end{figure}

The bars of the SMC and LMC have been surveyed with the OGLE and MACHO
surveys, and the resulting databases have been extensively data mined
for red variables, since the pioneering work by Wood (1999). By
selecting on pulsation amplitude different sequences are populated, as
illustrated in Figure~\ref{groen-PL-MC}, with the largest amplitude
Mira variables belonging to sequence ``C''.  For a summary of the
results of the micro lensing surveys on red variables in the MCs, I
refer to Groenewegen (2006b).

Works on other LG galaxies are those by:
Bersier \& Wood (2002) who describe 85 LPV candidates in Fornax.
Gallart et al. (2004) who propose 6 LPV candidates in Phoenix.
Rejkuba et al. (2003), Rejkuba (2004) who found 1146 LPVs in NGC 5128.
Snigula et al. (this conference), who identify
11 / 2 / 52 / 0  LPV candidates in, respectively,
Leo A / GR 8 / Pegasus / DDO 210.

\smallskip
A number of studies appeared recently on variables in M31: 
Ansari et al. (2004) present results from the AGAPE gravitational
micro lensing survey. They present astrometry and photometry on 1579
variables stars in a 10\arcmin x 14\arcmin\ field.  They observed the
field for 3 years and obtained between 40-80 epochs. The $(B-R)$ colours
and $R$ magnitudes suggest that the majority are LPVs. Periods are
only presented for 54 objects however.
Fliri et al. (2006) present results from the WeCAPP micro lensing survey. 
A field of 16\arcmin x 16\arcmin\ was monitored in $R$ and $I$ over 3
years with 200-400 epochs. 23781 variable sources are detected, of
which 19167 are classified as ``regular or semi-regular red variables''.
Mould et al. (2004) monitored 33 fields covering 5 stripes of about
10\arcmin x 60\arcmin\ in $I$ for over 6 years, with 13-17 epochs per
field.  Single-epoch $JHK$ photometry was also carried out, and
astrometry and 4-band photometry is presented for 1915 LPVs. Using the
period derived from the $I$-band monitoring and the single-epoch
$K$-band they present a period-luminosity relation. Naturally there is
quite some scatter, but more than one would expect even from
single-epoch data.  Some of the stars that are significantly brighter
than the LMC $PL$-relations shifted to the distance of M31 can be
identified with supergiants (as they seem associated with the ring of
star formation in M31), but there are also objects at periods longer
than 600 days that are 2 mag fainter than the $PL$-relation.

\subsection{ Narrow-band surveys}

This technique uses the specific spectral characteristic of late-type
stars, where strong molecular TiO bands develop in M-stars, and C$_2$
and CN bands in C-stars.  First introduced by Wing (1971) and Palmer
\& Wing (1982) and then applied by Richer et al. (1984) and Aaronson
et al. (1984) the method typically uses two broad-band filters from
the set $V,R,I$, and two narrow-band filters near 7800 and 8100 \AA,
which are centred on a CN-band in carbon stars (and near-continuum in
oxygen-rich stars), and a TiO band in oxygen-rich stars (and continuum
in C-stars), respectively. In an [78-81] versus $[V-I]$ (or $[R-I]$)
colour-colour plot, carbon stars and late-type oxygen-rich stars
clearly separate redwards of $(V-I) \approx$ 1.6.  For an illustration
of this, see Cook \& Aaronson (1989) or Nowotny \& Kerschbaum (2002).

At present a large fraction of LG galaxies have been
surveyed, at least partially, using these narrow-band filters. For
recent reviews see Azzopardi (1999) and Groenewegen (1999, 2002, 2006a,b). 
The most recent works {\it not listed in these reviews} are the surveys by:
Battinelli \& Demers (2005a) who find 15 C-stars in the disk of M31
beyond 30 kpc along it major axis, and Battinelli \& Demers (2006) who
identify 46 C-stars in DDO 190. In addition, Battinelli \& Demers
(2005b,c) summarise their work on over 10 LG galaxies regarding the
standard candle aspect of the C-star $I$-band luminosity function, and
the calibration of the C/M ratio versus metallicity.
Kerschbaum et al. (2004) report 51 C-stars in (the direction of) M32, and
Spindler et al. (2006) found 40 C-stars in Leo {\sc i} (19 new),
11 in Leo {\sc ii} (6 new), and 2 in Draco (no new).
Finally, Harbeck et al. (2005) found one C-star candidate in And {\sc ix}.

\begin{figure}[!ht]

\centering
\begin{minipage}{0.78\textwidth}
\resizebox{\hsize}{!}{\includegraphics{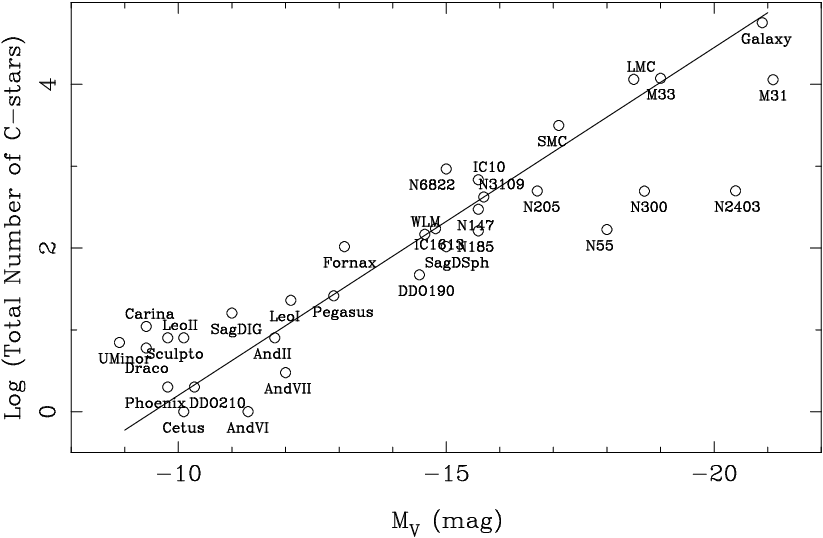}}
\end{minipage}


\caption{Updated version (cf. Groenewegen 2006a) of the relation
between the total number of C-stars and absolute (visual) magnitude of
the parent galaxy. The line is not a fit to the data but comes from
Battinelli \& Demers (2005c) who considered mainly the dozen
galaxies they have surveyed. }
\label{groen-C}
\end{figure}

\begin{figure}[!ht]

\centering
\begin{minipage}{0.78\textwidth}
\resizebox{\hsize}{!}{\includegraphics{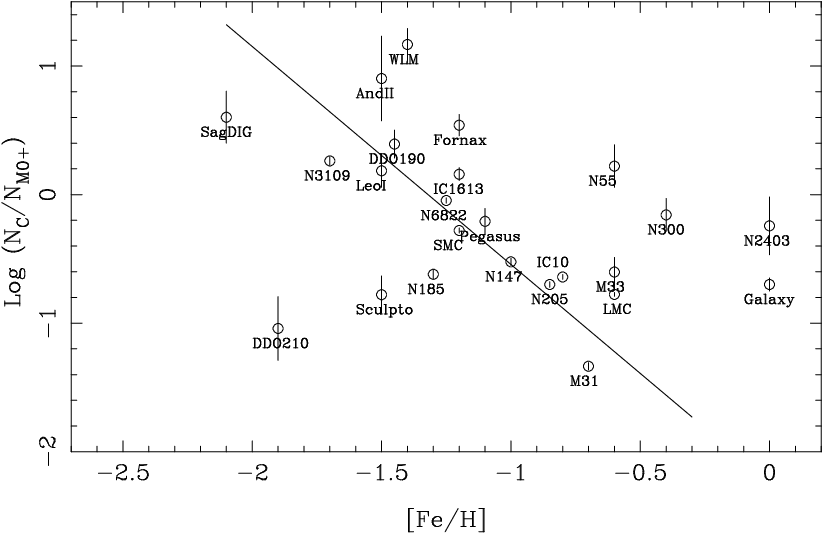}}
\end{minipage}


\caption{Updated version (cf. Groenewegen 2006a) of the relation
between the total number of C-stars and absolute (visual) magnitude of
the parent galaxy. The line is not a fit to the data but comes from
Battinelli \& Demers (2005b) who considered the dozen galaxies they
surveyed. }
\label{groen-CM}
\end{figure}

\smallskip
Figures~\ref{groen-C} and \ref{groen-CM} show updated versions
(cf. Groenewegen 2006a) of the well known relations between the total
number of C-stars and absolute (visual) magnitude of the parent
galaxy, and the relation between C-to-M number ratio and (mean)
metallicity of the parent galaxy.  The lines are not fits to the data but taken
directly from Battinelli \& Demers (2005b,c) who discussed specifically
the dozen galaxies they observed over the last 6 years. The extended
dataset fit these lines very well. Outliers are NGC 55, 300, 2403 whose
surveys for AGB stars from the 1980s are incomplete.

\begin{figure}[!ht]
\plotone{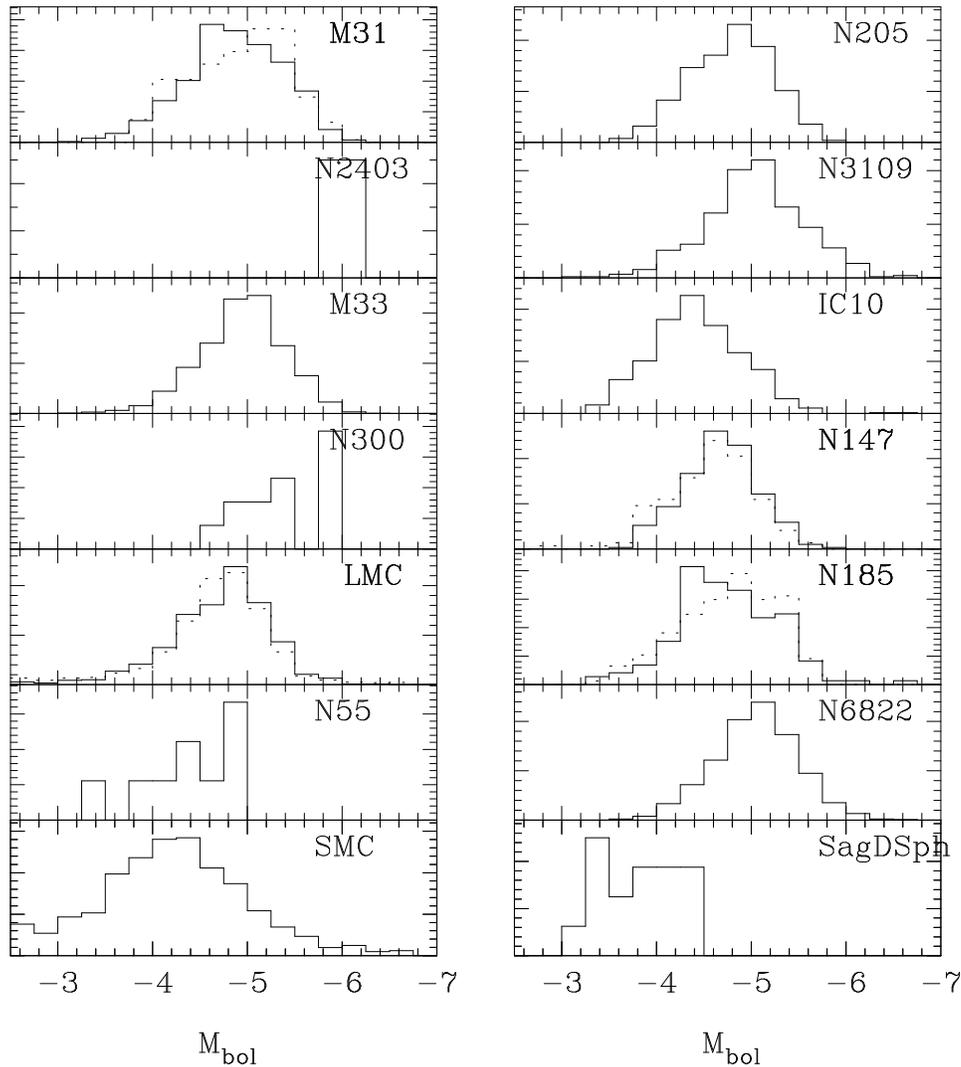}
\caption{Updated version (cf. Groenewegen 2006a) of the normalised
carbon star luminosity functions (LFs), ordered by decreasing $M_{\rm V}$ 
of the galaxies, ordered top to bottom, left to right.
The number of C-stars used to calculate the LFs varies from galaxy to
galaxy. In the case of SMC and LMC, the lowest luminosity bin is cumulative. 
}
\vspace{-3mm}
\label{groen-LF1}
\end{figure}

\begin{figure}[!ht]
\plotone{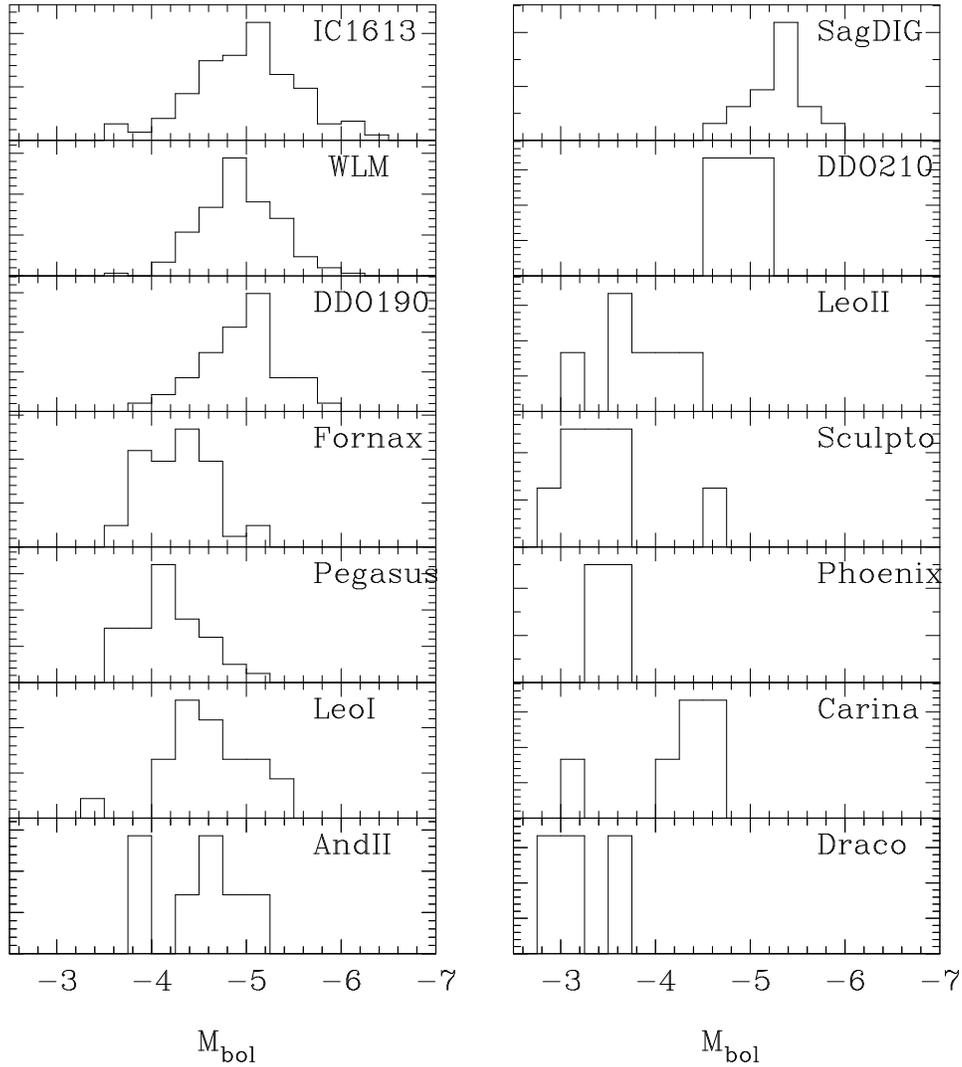}
\caption{Luminosity functions: continued }
\vspace{-3mm}
\label{groen-LF2}
\end{figure}

\smallskip
Figures~\ref{groen-LF1} and \ref{groen-LF2} show updated versions of
the carbon star luminosity functions.  Compared to Groenewegen
(2006a), (a) the $V,I$ data from Brewer et al. (1995) for M31 (the
dashed line) has been converted to $m_{\rm bol}$ using the bolometric
correction from Nowotny et al. (2003) [like was already the case for
the other galaxies in this plot for which $V,I$ data is available] and
the LF now agrees much better with the LF of Battinelli et al. (2003,
solid line), (b) M33, IC 10 and DDO 190 are added, and (c) the LF of
Ursa Minor is not reproduced again for reasons of space.

\smallskip
The fact that LG galaxies are increasingly surveyed in the infrared as
well, allows one to compare the selection of M- and C-stars using
infrared data with that using the narrow-band filter system.  Demers
et al. (2006) has done this for NGC 6822. AGB stars are selected as
being brighter than the tip of the RGB, with M-stars having $1.07 <
(J-K)_0 < 1.36$, and C-stars having $(J-K)_0 \ge 1.36$ (following
Cioni \& Habing 2005b). In the region of overlap the objects are
correlated with the narrow-band data of Letarte et al. (2002). From
the 85 infrared selected C-stars candidates, 69 fall in the region in
the [CN-TiO] versus $(R-I)$ colour-colour diagram occupied by the
C-stars, and 16 do not (10 are inside the region used to define
M-stars, and 6 show bluer colours). These stars have a $(J-K)$ colour
close to the limit used.
Of the 207 infrared selected M-star candidates, 21 are in or close to
the region defined by the C-stars, 3 have $(R-I)$ colours indicating a
cool object but have an [CN-TiO] index intermediate between C- and
M-stars, about 45 have bluer colours than used to define M-stars, and
about $\frac{2}{3}$ of the sample are actually inside the region that
define M-stars.

These results indicate that a slightly more ``purer'' sample of
infrared selected C-stars might have been obtained by selecting
slightly redder objects. It is not clear if this means redder in
$(J-K)_0$ than 1.36 or if the foreground extinction, which is
substantial with $A_{\rm V}$ = 0.8, was slightly underestimated.

It is not discussed in Demers et al. (2006) if increasing the lower
limit for M-star selection would remove the bluer stars in $(R-I)$. 
The current selection criterion implies an oxygen-rich sample that is
roughly 10\% contaminated by C-stars, and with 20\% stars that are
bluer than used to define M-stars in the narrow-band surveys.

A different view on the same topic is the following. Letarte et
al. (2002) identify 904 C-stars and derive a C/M ratio of 1.0 $\pm$ 0.2 
in a 28\arcmin\ $\times$ 42\arcmin\ centred on NGC 6822.
Cioni \& Habing (2005b) observed a 20\arcmin\ $\times$ 20\arcmin\ area
in $IJK$. Selecting stars above the TRGB and using 1.36 as the
division between C- and O-stars in $(J-K)$, they find 1511 C-stars and
a C/M ratio of 0.32.
Kang et al. (2006) observed a 6.3\arcmin\ $\times$ 3.6\arcmin\ field.
They use the criterion $(J-K) > 1.53$ and $(H-K) > 0.5$ to select 141
C-stars and derive a C/M ratio of 0.27. Their selection criterion is
based on the NIR colours of the known C-stars from Letarte et al.

The fact that Kang et al. assume a redder limit for the selection of
C-stars probably explains why they have relatively fewer C-stars
compared to Cioni \& Habing; 141 $\times$ (20 $\times$ 20) / (6.3
$\times$ 3.6) is about 2500 while Cioni \& Habing have 1511 C-stars (the
density of C-stars is not uniform as well). As the C/M ratio is very
similar, their selection based on $(J-K)$ and $(H-K)$ must also have
led to fewer M-stars than in Cioni \& Habing.

The large discrepancy in C/M ratio between the infrared works and the
narrow-band work is likely due to a larger number of ``M-stars'' in
the infrared works. As already mentioned earlier in discussing the
results by Demers et al., even selecting M-stars as having $1.07 <
(J-K) < 1.36$ results in more M-stars that found by the narrow-band
surveys. {\bf No} lower limit on the selection of M-stars in $(J-K)$,
as appears to be the case in Kang et al. and Cioni \& Habing (the bluest
stars included have $(J-K) = 0.8$), will therefore lead to too many
``M-stars'' and hence to a lower C/M ratio.
\vspace{-5mm}

\section{Conclusions and future work}  

The narrow band surveys of the LG are fairly complete. Notable
exceptions are Fornax DSph$^{\star}$ (at 140 pc distance), LGS3 (620
pc), Leo A$^{\star}$ (800 pc), Sextans B$^{\star}$ (1320 pc), 
Sextans A$^{\star}$ (1440 pc), IC 5152 (1700 pc), GR 8 (2200 pc). 
It might also be worthwhile to re-do NGC 55$^{\star}$ (2200 pc), NGC
300$^{\star}$ (2200 pc), NGC 2403$^{\star}$ (3600 pc), for which
incomplete data from the 1980's exist. A search for AGB stars would
certainly be successful as PNe are known to exist in almost all of
these galaxies (those marked by a $^{\star}$).  Crowding may be an
issue from the central parts of the most distant galaxies (see
e.g. DDO 190 at 2.8 Mpc), but is not an issue in e.g. NGC 3109
at 1.3 Mpc.

The largest telescopes which currently have the CN, TiO narrow-band
filter system installed are SOAR (4.1m, 5.2\arcmin\ FoV), CFHT (3.6m,
42$^\prime$ x 28$^\prime$ FoV), WIYN (3.5m, 9.6$^\prime$ FoV) and the TNG
(3.5m, 9.4$^\prime$ FoV).

\medskip
The first results of the IRAC {\it Spitzer} observations of WLM
presented at this conference reveal a significant population of red
stars not detected in the narrow-band survey for this galaxy. In the
near future, analysis of IRAC observations of other LG galaxies will
reveal if this is a more general phenomenon. We may need to revise our
understanding of mass loss at low metallicity if the presence of
obscured AGB stars is more widespread that we now believe.

\medskip
Related to this, an increasing number of LG galaxies is being observed
in the near-infrared. Simple infrared criteria have been used to
separate C- from M-stars but a comparison to the results obtained
using the narrow-band filter system suggests that the infrared
criteria need refinement as they lead to the selection of too many C,
and way too many oxygen-rich stars that are earlier than spectral type
M (as defined by the narrow-band surveys).

\medskip
Additional information that might be used in classification is
variability.  Projects like super-MACHO, OGLE-{\sc iii} continue to
observe the MC and Galactic Bulge region.  Some work on LG galaxies is
being carried out using the SIRIUS camera on the IRSF in
South-Africa. A revolution is this respect in the more distant future
will be the LSST (see Ivezi\'c, this conference) that will observe the
visible sky in a few days only.
 
\medskip
In a nearer future abundance studies of AGB stars in LG galaxies will
become available. Some results have been presented by de Laverny et
al. (2006) for 1 SMC and 2 C-stars in the Sag DSph using VLT/UVES, and
by Wahlin et al. (2006, and this conference) who have observed 50
C-stars in the MCs, Sculptor, Carina and Fornax using VLT/ISAAC. With
the commissioning of VLT/CRIRES a very-resolution infrared spectrograph
will become available that is very well suited for abundance studies.

\medskip
Observations that also will be significant in an intermediate future
are heterodyne observations with ALMA. For a few dozen bright and red
AGB stars it will be possible to detect the lower-transition CO lines
(Groenewegen 1996) and thereby determine the expansion velocity and
the gas mass loss rate, and by comparison to the dust mass loss rate,
the important dust-to-gas ratio.
\vspace{-3mm}



\end{document}